# Multi-agent Coordination in Directed Moving Neighborhood Random Networks

Yilun Shang[1]

**Abstract**

In this paper, we consider the consensus problem of dynamical multiple agents that communicate via a directed moving neighborhood random network. Each agent performs random walk on a weighted directed network. Agents interact with each other through random unidirectional information flow when they coincide in the underlying network at a given instant. For such a framework, we present sufficient conditions for almost sure asymptotic consensus. Some existed consensus schemes are shown to be reduced versions of the current model.

## 1. Introduction

Synchronization and consensus problems in multi-agent systems have a long history [3]. Over the last few years, considerable research attention has been paid to cooperative group coordination, which has broad applications in a variety of areas including vehicle formation, distributed sensor networks, coupled oscillator systems, animal aggregation and epidemics; see the survey papers [10, 14, 19] and references therein. In a consensus problem, a group of autonomous agents seeks to agree on some quantities of interest via a process of distributed decision making based upon local information interactions. A consensus is reached when all agents in the system tend to attain agreement on the quantities of interest as time evolves, that is, they converge to a common value.

The ability to achieve coherent behavior (or consensus) of multi-agent system is intimately related with the coupling topology of the underlying communication graph associated with the agents' positions. For example, Ali Jadbabaie et al. [6] analyzed a simplified model of flocking by Vicsek et al. [20] showing that all agents will synchronize eventually, provided the communication graph switching deterministically over time is periodically jointly connected. More recently, there has been some interest in stochastic consensus problems. [4] deals with the agreement problem among a group of static agents with the communication graph presented as Erdös-Rényi random graph. The results are further extended by [12] to include weighted and directed information flows. In [18], the authors reveal this problem can be reduced to weak ergodicity of a sequence of stochastic matrices. However, the agents are static in these works, in other words, the switching of communication graph is merely due to the adjunctive random mechanism rather than the motion of agents.

The moving neighborhood network proposed in [16] is meant to treat stochastic consensus among dynamical agents (in social interaction and epidemics originally), where each agent carries an oscillator and diffuses in the environment. The computer simulation in the aforementioned paper shows that synchronization is possible even when the communication network is spatially disconnected in general at any frozen time instant. This is in sharp contrast with the behavior of Vicsek model mentioned above. Analytical formalizations are provided by authors of [17, 13, 9] with different consensus protocols (continuous-time with dwell time or discrete-time) through fast switching techniques [17] and stochastic stability theory [7]. In particular, they specify the moving neighborhood network model as a group of $n$ identical agents implementing simple random walks on a fixed finite connected graph. The vertices of the moving neighborhood network are represented by the agents, and the edges are determined by their locations in the underlying graph, i.e. a link between two agents appears if and only if they reside in the same node simultaneously. The time evolution of the moving neighborhood graph governed by the random walk of agents is called network dynamics and the time evolution of the corresponding oscillators system is called system dynamics, which is coupled together with the network dynamics.

[1]Department of Mathematics, Shanghai Jiao Tong University, Shanghai 200240, CHINA. email: shyl@sjtu.edu.cn

The purpose of this note is to generalize the above results and go a further step in the direction of [9]. Our extension is threefold. Firstly, we consider the underlying fixed graph as a weighted directed graph. In other words, random walks on weighted directed graph is tackled. Secondly, further randomness is involved in the moving neighborhood graph. A probability is associated (not necessarily independent) to the link between two agents when they occupy the same node, which can be interpreted as communication links may be unreliable due to disturbance. We allow different probabilities and unidirectional communications. Therefore the topology of the moving neighborhood graph is no longer strictly separated clique-like. Finally, we study a discrete-time consensus protocol with inhomogeneous weights, which quantify the way the agents influence each other. By these modifications, some pertinent known results can be recovered (see the remarks in Section 2 & 4), and a number of introduced parameters enhance more flexibility.

The rest of this paper is organized as follows. In Section 2, we elaborate on our formal model under consideration. Section 3 contains some mathematical preliminaries and the statement of main result. Proofs are given in Section 4.

## 2. Problem setup

Let $\bar{G} = (\bar{V}, \bar{E}, W)$ be a weighted digraph with vertex set $\bar{V} = [m]$, here $[m] := \{1, 2, \cdots, m\}$. The elements in $\bar{V}$ will be referred to as nodes. $\bar{E}$ is a set of ordered pair of elements of $\bar{V}$ called arcs. $W = (w_{ij})$ is the weight matrix, where $w_{ij} > 0$ if $(i, j) \in \bar{E}$, and $w_{ij} = 0$ otherwise. Notice that we do not exclude loops. We impose the following condition on the graph $\bar{G}$.

**Assumption 1.** The underlying graph $\bar{G}$ is strongly connected and the gcd (greatest common divisor) of all cycle lengths in $\bar{G}$ is 1.

This assumption will be elucidated in the next section. Let $t \in \mathbb{N} \cup \{0\}$ be the discrete time step. We consider $n$ identical agents signified by $\{v_1, v_2, \cdots, v_n\}$ as random walkers meandering on $\bar{G}$, moving randomly to a neighbor of their current location in $\bar{G}$ at each time step. For each agent, the random neighbor that is chosen is not affected by the agent's previous trajectory. The $n$ random walk processes are mutually independent. A time-varying moving neighborhood graph, $G = (V, E, P)$, is constructed as follows. Let $V = V(t) = \{v_1, v_2, \cdots, v_n\}$ be the vertex set. $P = (p_{ij})$ is an $n \times n$ matrix with entries $0 < p_{ij} \leq 1$ for $i \neq j$, and $p_{ii} = 0$ for $1 \leq i \leq n$. $P$ is dubbed as the linkage probability matrix. Given $v_i$ and $v_j$ meet at the same node simultaneously, an arc $(v_i, v_j)$ originating from $v_i$ and terminating in $v_j$ appears with probability $p_{ij}$. The arc set is denoted by $E = E(t)$. We emphasize here that the selections of arcs are independent with the random walk processes, but are not required to be independent with each other. By definition, $G = G(t)$ has no loops with probability 1.

Let $X_i(t) \in \mathbb{R}$ be the state (attitude, heading, opinion etc.) of agent $v_i$ at time t. The consensus protocol can be expressed as

$$X_i(t+1) = X_i(t) + \varepsilon \sum_{j \in N_i(t)} b_{ij}(X_j(t) - X_i(t)) \tag{1}$$

where $N_i(t)$ is the index set of neighbors of agent $v_i$ at instant t, i.e. $N_i(t) := \{j|(v_i, v_j) \in E(t)\}$. Here, weighting factor $b_{ij} > 0$ for $i \neq j$, and $b_{ii} = 0$ for $1 \leq i \leq n$. Let $A(t) = (a_{ij}(t))$ be the adjacency matrix of the moving neighborhood graph $G(t)$, whose entries are given by,

$$a_{ij}(t) = \begin{cases} b_{ij}, & (v_i, v_j) \in E(t) \\ 0, & \text{otherwise} \end{cases}$$

for $1 \leq i, j \leq n$. Moreover, we assume the following

**Assumption 2.** Suppose that $\sum_{j=1}^{n} a_{ij}(t) = \sum_{j=1}^{n} a_{ji}(t)$, for $1 \leq i \leq n$.

The implication of Assumption 2 is the out-degrees are equal to in-degrees for every $v \in V(t)$. A similar definition used in [11] is referred to as "balanced" digraph. Note that in the context of [11], the underlying communication network is static, while in the current case, we dictate each trajectory of $G(t)$ should be "balanced". However, this requirement is not very stringent after all. For example, if we postulate $G(t)$ can only have bidirectional (or undirected) edges, that is, $b_{ij} = b_{ji}$ and $p_{ij} = p_{ji}$ for $1 \leq i, j \leq n$, then Assumption 2 is satisfied automatically. Also, if $b_{ij} \equiv b$ for $i \neq j$ and suppose for each $v \in V(t)$, the numbers of arcs going out and coming in are equal, then Assumption 2 is filled, too. It is worth noting that since the selections of arcs can be arbitrarily dependent with each other as mentioned before, the latter of the two examples above can actually occur.

Next, let $\triangle := \max_{1 \leq i \leq n}(\sum_{j=1}^{n} b_{ij})$, and we further assume $\varepsilon \in (0, 1/\triangle)$. This assumption is natural; see Section 4.

As usual, denote the linear subspace $\mathcal{M} := \{x = (x_1, x_2, \cdots, x_n)^T \in \mathbb{R}^n | x_i = x_j \ \forall \ 1 \leq i, j \leq n\}$ as the synchronization manifold. A consensus is said to be reached if the states of agents converge into $\mathcal{M}$ as $t$ approaches infinity. We aim to show that the $n$ agents in $V$ reach consensus almost surely.

We end this section by two remarks.

**Remark 1.** Suppose that the linkage probabilities $p_{ij}(t)$ and the weighting factors $b_{ij}(t)$ are time-dependent, then the main result in this paper holds verbatim as long as $p_{ij}(t)$ and $b_{ij}(t)$ converge.

**Remark 2.** If we take the underlying graph $\bar{G}$ to be a single node, then our framework somewhat reduces to that of [4, 18, 12].

## 3. Preliminaries and main result

Let $\{Y_i(t), t \in \mathbb{N} \cup 0\}$ denote the random walk process performed by agent $v_i$ such that $Y_i(t) \in \bar{V}$ designates the position of $v_i$ in the underlying graph $\bar{G}$ at instant $t$. For $1 \leq i \leq n$, let $\pi_i(t) = (\pi_{i1}(t), \cdots, \pi_{im}(t))^T$ be the probability distribution of $Y_i(t)$ at time $t$, that is, $P(Y_i(t) = j) = \pi_{ij}(t)$ for $j \in \bar{V}$. By the construction, $\{Y_i(t), 1 \leq i \leq n\}$ are $n$ independent homogeneous Markov chains with finite state space $\bar{V}$, sharing the same transition probability matrix $Q = (q_{ij})$, whose entry is given by

$$q_{ij} = \frac{w_{ij}}{d_i} \quad \text{for } 1 \leq i, j \leq m \tag{2}$$

where $d_i = \sum_{j=1}^{m} w_{ij}$ represents the out-degree of node $i$. Also recall that $q_{ij} = P(Y_1(t+1) = j | Y_1(t) = i)$. Notice that by Assumption 1 $d_i \neq 0$, so (2) is well-defined.

As is known, a finite Markov chain converges to a unique stationary distribution if it is ergodic, i.e. irreducible and aperiodic (see e.g. [15] Thm. 4.2). It is clear that the transition probability matrix $Q$ is irreducible if and only if $\bar{G}$ is strongly connected. Also, it's easy to see that the gcd of all cycle lengths in $\bar{G}$ is 1 if and only if all eigenvalues of $Q$ other than 1 have modules strictly less than 1. Actually, the above assertions may be proven by employing Perron-Frobenius Theorem [15] and some other equivalent conditions in terms of the ergodicity of random walk are given in [2]. Consequently, in view of Assumption 1, $Y_i(t)$ converges to a unique stationary distribution $\pi = (\pi_1, \cdots, \pi_m)^T$ for $1 \leq i \leq n$, that is, $\pi^T = \pi^T Q$. For instance, if $\sum_{j=1}^{m} w_{ij} = \sum_{j=1}^{m} w_{ji}$ for $1 \leq i \leq m$, then we have $\pi_i = d_i / \sum_{i,j=1}^{m} w_{ij}$ for $1 \leq i \leq m$, see e.g. [8]. As for the rate of convergence, we have the following lemma, (see [1] Thm. 8.9 for a proof).

**Lemma 1.** ([1]) *For a finite ergodic Markov chain with transition probability matrix $P$, let $p_{ji}^{(t)}$ be the $(j, i)$ entry of $P^t$, $t \in \mathbb{N} \cup 0$, then there is a unique stationary distribution $\{p_i\}$ and*
$$|p_{ji}^{(t)} - p_i| \leq \lambda \rho^t,$$
*where $\lambda \geq 0$, $0 \leq \rho < 1$.*

Next, let $D(t) = \text{diag}(d_{11}(t), \cdots, d_{nn}(t))$ be the out-degree diagonal matrix of the moving neighborhood graph $G(t)$. To be precise, $d_{ii}(t) = \sum_{j=1}^{n} a_{ij}(t)$, for $1 \leq i \leq n$, where $a_{ij}(t)$ are the entries of $A(t)$ defined in Section 2. Notice that $A(t)$ and $D(t)$ are non-negative time-dependent random matrices. Since $a_{ij}(t)$ is a two-point distribution, its expectation is shown to be given by

$$E(a_{ij}(t)) = b_{ij} p_{ij} \sum_{k=1}^{m} \pi_{ik}(t) \pi_{jk}(t), \quad \text{for } i \neq j, \tag{3}$$

and $a_{ii}(t) = 0$. The diagonal entries of $D(t)$ can be written as

$$d_{ii}(t) = \sum_{k=1}^{m} \sum_{j=1, j\neq i}^{n} b_{ij} 1_{[Y_i(t)=k, Y_j(t)=k, (v_i, v_j) \in E(t)]}.$$

Thereby, $E(d_{ii}(t)) = \sum_{k=1}^{m} \sum_{j=1}^{n} b_{ij} p_{ij} \pi_{ik}(t) \pi_{jk}(t)$, for $1 \leq i \leq n$. Now let $L(t) = D(t) - A(t)$ denote the Laplacian matrix [2] of $G(t)$, and it's expectation at time t is $EL(t) = ED(t) - EA(t)$.

For a sequence of random elements $Z(t)$, the ergodic limit, $E^*(Z)$, is defined by $E^*(Z) = \lim_{t \to \infty} E(Z(t))$, whenever the limit exists. $E^*(Z)$ describes the long-run average respect to its stationary distribution [13]. By the above discussion, the ergodic limit of (3) is $E^*(a_{ij}) = b_{ij} p_{ij} \sum_{k=1}^{m} \pi_k^2$ for $i \neq j$ and $E^*(a_{ii}) = 0$ for $1 \leq i \leq n$. Likewise, $E^*(d_{ii}) = \sum_{k=1}^{m} \pi_k^2 \sum_{j=1}^{n} b_{ij} p_{ij}$ for $1 \leq i \leq n$.

Now we make a reference to Schur product of two real matrices, which is also known as Hadamard product [5]. Let $C = (c_{ij})$, $E = (e_{ij})$ be two $n \times m$ real matrices, then the Schur product, $C \circ E = (c_{ij} e_{ij})$, is simply the product of corresponding entries of $C$ and $E$. Obviously, $C \circ E = E \circ C$. Utilizing this notation, we obtain $E^*(A) = \pi^T \pi B \circ P$ and $E^*(D) = \pi^T \pi \text{diag}(\sum_{j=1}^{n} b_{1j} p_{1j}, \cdots, \sum_{j=1}^{n} b_{nj} p_{nj})$. Hereby, we have

$$E^*(L) = \pi^T \pi \left( \text{diag}\left( \sum_{j=1}^{n} b_{1j} p_{1j}, \cdots, \sum_{j=1}^{n} b_{nj} p_{nj} \right) - B \circ P \right). \tag{4}$$

$E^*(L)$ is positive semi-definite and have an eigenvalue 0 since $E^*(L)1 = 0$, where 1 is the all-1 column vector. We need the following lemma regarding Schur product.

**Lemma 2.** *Suppose $C$, $E$ are two $n \times n$ real matrices and $x$, $y$ are two $n \times 1$ real vectors, then*
$$y^T(C \circ E)x = \text{tr}\big(\text{diag}(y^T)^T C \text{diag}(x^T) E^T\big).$$

The proof will be given in the next section. An asymptotic stability result in [7] is restated below for convenience, which is an analogy of the deterministic Lyapunov stability theorem.

**Lemma 3.**([7], pp.195) *Let $\{X_n\}$ be a Markov chain on state space $S$. Suppose that there is a non-negative function $\xi(x)$ satisfying*
$$E(\xi(X_1)|X_0 = x) - \xi(x) = -\eta(x),$$
*where $\eta(x) \geq 0$ on the set $Q_\beta := \{x : \xi(x) < \beta\}$. Then*
$$P\big(\sup_{n \geq 0} \xi(X_n) \geq \beta | X_0 = x\big) \leq \xi(x)/\beta,$$
*and accordingly, $\eta(X_n) \to 0$ almost surely for paths which remain in $Q_\beta$.*

It is at this stage, we state our main result as follows.

**Theorem 1.** *Under the circumstances and assumptions presented above, the stochastic system expressed by (1) reaches consensus almost surely.*

## 4. Proofs

This section includes the proofs of Lemma 2 and Theorem 1.

**Proof of Lemma 2.** Let $1 = (1, 1, \cdots, 1)^T$ be the $n \times 1$ vector. Note that $\text{diag}(y^T)1 = y$. Therefore, we get
$$\begin{aligned} y^T(C \circ E)x &= 1^T \text{diag}(y^T)^T (C \circ E)x \\ &= 1^T (\text{diag}(y^T)^T C \circ E)x \\ &= \text{tr}\big(\text{diag}(y^T)^T C \text{diag}(x^T) E^T\big) \end{aligned}$$

We have exploited a basic property: $(A\text{diag}(x^T)B^T)(i,i) = ((A \circ B)x)(i)$ in the last equality above. Here notation $A(i,i)$ means the $(i,i)$ entry of matrix $A$, and $x(i)$ the $i$th element of vector $x$. $\square$

**Proof of Theorem 1.** We rewrite the protocol (1) in a compact matrix form as
$$X(t+1) = F(t)X(t), \qquad (5)$$
where $X(t) = (X_1(t), \cdots, X_n(t))^T$ is the state vector of agents at instant $t$, and $F(t) = I_n - \varepsilon L(t)$. The spectrum of $F(t)$ satisfies $1 = \lambda_n(F(t)) \geq \cdots \geq \lambda_1(F(t)) \geq 1 - 2\varepsilon\triangle$, since the eigenvalues of Lapacian $L(t)$ are $2\triangle \geq \lambda_n(L(t)) \geq \cdots \geq \lambda_1(L(t)) = 0$; see [2] for more about Laplacian spectrum.

Recall we assume $\varepsilon \in (0, 1/\triangle)$ in Section 2, which yields $|\lambda_i(F(t))| < 1$ for $1 \leq i < n$ and causes the state transition matrix $F(t)$ to be stable at each time step, (see e.g. [10] Lemma 3). By the Assumption 2 and iterative equation (5), we obtain
$$1^T X(t+1) = 1^T F(t) X(t) = (F(t)X(t))^T 1 = X(t)^T F(t)^T 1 = (X(t)^T 1)^T = 1^T X(t).$$

From this, it is clear that the projection of the state vector $X(t)$ on synchronization manifold $\mathcal{M}$ is a constant. In fact, let $\alpha := \frac{1}{n}\sum_{i=1}^{n} X_i(0)$, then $1^T X(t) = n\alpha$ for all $t$.

Now we may decompose $X(t)$ as

$$X(t) = X^{pa}(t) + X^{pe}(t), \quad (6)$$

where $X^{pa}(t) \in \mathcal{M}$ and $X^{pe}(t) \perp \mathcal{M}$. The superscript "pa" stands for "parallel", while "pe" for "perpendicular". We thereby get $||X^{pa}(t)|| = \frac{1}{\sqrt{n}} 1^T X(t) = \sqrt{n}\alpha$. Here and in the sequel, we take $||\cdot||$ as 2-norm for vectors and induced 2-norm for matrices [5]. The disagreement among the agents now can be described by $||X^{pe}(t)||^2 = ||X(t)||^2 - n\alpha^2$. To prove the theorem, it suffices to show $||X^{pe}(t)|| \to 0$ almost surely as $t \to \infty$.

We follow [9] to introduce a Lyapunov function $\xi(x) = x^T x - \alpha^2 n$, and let

$$\xi(t) := \xi(X(t)) = X(t)^T X(t) - \alpha^2 n.$$

Our plan is to use Lemma 3 to show $\xi(t) \to 0$ almost surely, as $t \to \infty$. Note that $X(t)$ is indeed a Markov chain. With this in mind, we evaluate by employing (5),

$$E(\xi(t+1)|X(t)) = X(t)^T E(F(t)^T F(t)) X(t) - \alpha^2 n.$$

Therefore we get

$$E(\xi(t+1)|X(t)) - \xi(t) = X(t)^T EH(t) X(t) \quad (7)$$

where, $H(t) := F(t)^T F(t) - I = \varepsilon^2 L(t)^T L(t) - \varepsilon(L(t)^T + L(t))$.

Let $L_i$ be the $i$th possible realization of $L(t)$, with probability $p_i(t)$ at instant $t$, and denote $\Omega$ as the collection of all $L_i$. We hereby may write the expectation of $L(t)$ as $EL(t) = \sum_{i=1}^{|\Omega|} L_i p_i(t)$. $|\Omega|$ is the cardinality of $\Omega$. Since $L(t)$ inherited the property of the random walks of the agents, it is also an ergodic Markov chain with state space $\Omega$. Let $p_i$ be the $i$th component of the unique stationary probability distribution. Then we may apply Lemma 1 and interpret the corresponding meaning of notations in the current situation. By doing so, we derive

$$\begin{aligned} |p_i(t) - p_i| &= \left|\sum_{j=1}^{|\Omega|} p_{ji}^{(t)} p_j(0) - p_i\right| = \left|\sum_{j=1}^{|\Omega|} (p_{ji}^{(t)} - p_i) p_j(0)\right| \\ &\leq \sum_{j=1}^{|\Omega|} |p_{ji}^{(t)} - p_i| p_j(0) \leq \lambda \rho^t \sum_{j=1}^{|\Omega|} p_j(0) = \lambda \rho^t. \end{aligned} \quad (8)$$

Now going back to the equation (7), we have

$$X(t)^T EH(t) X(t) = \sum_{i=1}^{|\Omega|} p_i(t) X(t)^T H_i X(t) \quad (9)$$

where $H_i = \varepsilon^2 L_i^T L_i - \varepsilon(L_i^T + L_i)$ is the $i$th possible realization of $H(t)$ and is negative semi-definite with eigenvalues $0 = \lambda_n(H_i) \geq \cdots \geq \lambda_1(H_i)$. Using the decomposition (6), we get

$$\begin{aligned} X(t)^T H_i X(t) &= X^{pe}(t)^T H_i X^{pe}(t) + X^{pa}(t)^T H_i X^{pe}(t) \\ &\quad + X^{pe}(t)^T H_i X^{pa}(t) + X^{pa}(t)^T H_i X^{pa}(t) \end{aligned}$$

There are four terms on the right hand side of the above equation, the last three of which are actually zero. To see why let's take the second term as an example,

$$\begin{aligned} X^{pa}(t)^T H_i X^{pe}(t) &= \varepsilon^2 X^{pa}(t)^T L_i^T L_i X^{pe}(t) - \varepsilon X^{pa}(t)^T L_i^T X^{pe}(t) - \varepsilon X^{pa}(t)^T L_i X^{pe}(t) \\ &= \varepsilon^2 (L_i X^{pa}(t))^T L_i X^{pe}(t) - \varepsilon (L_i X^{pa}(t))^T X^{pe}(t) \\ &\quad - \varepsilon (L_i^T X^{pa}(t))^T X^{pe}(t) \\ &= 0 \end{aligned}$$

since $L_i X^{pa}(t) = L_i^T X^{pa}(t) = 0$. Thereby we have $X(t)^T H_i X(t) = X^{pe}(t)^T H_i X^{pe}(t)$. Plugging this equation into (9), we get

$$X(t)^T E H(t) X(t) = X^{pe}(t)^T \Big( \sum_{i=1}^{|\Omega|} p_i(t) H_i \Big) X^{pe}(t) = X^{pe}(t)^T E H(t) X^{pe}(t).$$

We have the ergodic limit, defined in Section 3, of $H(t)$ as

$$\begin{aligned} E^* H &= \lim_{t \to \infty} E H(t) = \lim_{t \to \infty} E\big(\varepsilon^2 L(t)^T L(t) - \varepsilon (L(t)^T + L(t))\big) \\ &= \varepsilon^2 E^*(L^T L) - \varepsilon (E^* L^T + E^* L) \end{aligned} \quad (10)$$

since we know $L(t)$ is an ergodic Markov chain.

Take $R(t) := EH(t) - E^* H$ as the remainder matrix, then we have

$$X^{pe}(t)^T E H(t) X^{pe}(t) = X^{pe}(t)^T (E^* H) X^{pe}(t) + X^{pe}(t)^T R(t) X^{pe}(t). \quad (11)$$

The second term on the right hand side of (11) can be calculated, by noting (8), as follows

$$\begin{aligned} X^{pe}(t)^T R(t) X^{pe}(t) &= ||X^{pe}(t)^T R(t) X^{pe}(t)|| = \Big|\Big|X^{pe}(t)^T \Big(\sum_{i=1}^{|\Omega|} H_i(p_i(t) - p_i)\Big) X^{pe}(t)\Big|\Big| \\ &\leq ||X^{pe}(t)||^2 \cdot \Big|\Big|\sum_{i=1}^{|\Omega|} H_i(p_i(t) - p_i)\Big|\Big| \\ &\leq \xi(t) |\Omega| \gamma \lambda \rho^t, \end{aligned} \quad (12)$$

where $\gamma := \max_{1 \leq i \leq |\Omega|} ||H_i||$. Wherefore, (12) tends to 0, since $0 \leq \rho < 1$ and $\xi(t)$ is bounded by (1).

Next, we turn to the first term on the RHS of (11). In view of (10),

$$\begin{aligned} X^{pe}(t)^T (E^* H) X^{pe}(t) &= \varepsilon^2 X^{pe}(t)^T (E^* L^T)(E^* L) X^{pe}(t) - \varepsilon X^{pe}(t)^T (E^* L^T) X^{pe}(t) \\ &\quad - \varepsilon X^{pe}(t)^T (E^* L) X^{pe}(t). \end{aligned} \quad (13)$$

By using Lemma 2, we have

$$\begin{aligned} X^{pe}(t)^T (B^T \circ P^T) X^{pe}(t) &= \mathrm{tr}\big(\mathrm{diag}(X^{pe}(t)^T) B^T \mathrm{diag}(X^{pe}(t)^T) P\big) \\ &= \mathrm{tr}\big(\mathrm{diag}(X^{pe}(t)^T) B \mathrm{diag}(X^{pe}(t)^T) P^T\big) \\ &= X^{pe}(t)^T (B \circ P) X^{pe}(t). \end{aligned}$$

Then it's easy to see the last two terms on the RHS of (13) is equal, by noting equation (4). Thereby we obtain

$$\begin{aligned}
X^{pe}(t)^T(E^*H)X^{pe}(t) &= \varepsilon^2 X^{pe}(t)^T(E^*L)^T(E^*L)X^{pe}(t) - 2\varepsilon X^{pe}(t)^T(E^*L)X^{pe}(t) \\
&\leq (2\triangle\varepsilon^2 - 2\varepsilon)X^{pe}(t)^T(E^*L)X^{pe}(t) \\
&= 2\varepsilon\pi^T\pi(\varepsilon\triangle - 1)X^{pe}(t)^T \\
&\quad \cdot \big(\text{diag}\big(\sum_{j=1}^n b_{1j}p_{1j}, \cdots, \sum_{j=1}^n b_{nj}p_{nj}\big) - B \circ P\big)X^{pe}(t) \\
&\leq 2\varepsilon\pi^T\pi\triangle(\varepsilon\triangle - 1)\xi(t) \qquad (14)
\end{aligned}$$

The second inequality above holds because there is some $N(B) > 0$ depends only on B, such that $(E^*L)^T(E^*L) \leq 2\triangle(E^*L) + N(B)11^T$ and

$$X^{pe}(t)^T N(B)11^T X^{pe}(t) = N(B)(1^T X^{pe}(t))^T(1^T X^{pe}(t)) = 0.$$

The fourth inequality holds by noting that $\varepsilon\triangle - 1 < 0$ and

$$\big(\text{diag}\big(\sum_{j=1}^n b_{1j}p_{1j}, \cdots, \sum_{j=1}^n b_{nj}p_{nj}\big) - B \circ P\big) \geq \triangle I - M(B)11^T$$

for some $M(B) > 0$ depending only on $B$. The similar argument as above may be applied. Combing (12) and (14), we finally get

$$\begin{aligned}
E(\xi(t+1)|X(t)) - \xi(t) &\leq \xi(t)\big(|\Omega|\gamma\lambda\rho^t + 2\varepsilon\pi^T\pi(\varepsilon\triangle - 1)\triangle\big) \\
&:= \xi(t)\delta(t) := -\eta(t)
\end{aligned}$$

There exists $t_0 \in \mathbb{N}$ such that $\delta(t) < 0$ for $t \geq t_0$. Furthermore, $\delta(t) \to 2\varepsilon\pi^T\pi(\varepsilon\triangle - 1)\triangle$, as $t \to \infty$. Also, note that $\xi(t)$ is a Markov chain, since $X(t)$ is. By applying Lemma 3, where $\beta$ can take any positive number, we obtain

$$\eta(t) \to 0 \quad \text{as} \quad t \to \infty$$

almost surely, which yields $\xi(t) \to 0$ almost surely, as $t \to \infty$. The proof now is completed. □

**Remark 3.** Theorem 1 shows an average consensus which may be regarded as a kind of stochastic version of Theorem 9 in [11].